\documentclass[a4,aps,amsmath]{revtex4}
\usepackage{graphicx}
\newcommand{\be}{\begin{equation}}
\newcommand{\ee}{\end{equation}}
\newcommand{\ba}{\begin{array}}
\newcommand{\ea}{\end{array}}
\newcommand{\bea}{\begin{eqnarray}}
\newcommand{\eea}{\end{eqnarray}}
\newcommand{\bdm}{\begin{displaymath}}
\newcommand{\edm}{\end{displaymath}}

\begin{document}

\title{Dynamics of bootstrap percolation}

\author{Prabodh Shukla}
\email{shukla@nehu.ac.in}
\affiliation{%
Physics Department \\ North Eastern Hill University \\ 
Shillong-793 022, India}%

\begin{abstract}

Bootstrap percolation transition may be first order or second
order, or it may have a mixed character where a first order drop in the
order parameter is preceded by critical fluctuations. Recent studies have
indicated that the mixed transition is characterized by power law
avalanches, while the continuous transition is characterized by truncated
avalanches in a related sequential bootstrap process. We explain this
behavior on the basis of a through analytical and numerical study of the
avalanche distributions on a Bethe lattice. 

\end{abstract}

\maketitle

\section{Introduction}

Bootstrap percolation was introduced in 1979~\cite{chalupa} to study the
diminution and eventual destruction of magnetic order by non-magnetic
impurities in a magnet. As impurity increases, not only the concentration
of magnetic ions decreases but also an increasing number of magnetic ions
find themselves surrounded by insufficient number of magnetic neighbors to
retain their magnetism.  This effect accelerates the destruction of
magnetic order, and sometimes changes a continuous second order transition
as seen in ordinary percolation~\cite{fisher} to a discontinuous first
order transition. The first order transition encountered in bootstrap
problems has often a mixed character in the sense that the discontinuous
drop in magnetization is preceded by critical fluctuations characteristic
of a second order transition. The precise criterion for the occurrence of
a mixed transition is not very clear, and has been the subject of research
in recent years ~\cite{manna,farrow1,farrow2,parisi}.  The investigations
have relied upon numerical simulations in the case of regular
d-dimensional lattices, and analytic solutions in the case of a Bethe
lattice.  In some borderline cases, the presence of critical fluctuations
at the threshold of a discontinuity makes it difficult to distinguish
numerically between a true discontinuity, and a steep but continuous
decrease in the order parameter. A recent study of the distribution of
avalanches in the bootstrap dynamics has suggested a convenient criterion
for distinguishing between the two cases~\cite{farrow2}. The suggestion is
that the mixed transitions are characterized by power-law avalanches while
second order transitions are characterized by truncated avalanches. The
purpose of this article is to expand on this point by presenting a through
analytic and numerical study of the problem on a Bethe lattice. The Bethe
lattice does not capture all the complexities of bootstrap dynamics on
periodic lattices but it does provide useful insight into what makes the
avalanche distributions different in the two cases. The following
presentation is self-contained and does not assume any prior knowledge of
bootstrap percolation on the part of an interested reader.  However,
before we begin it may be appropriate to mention that in recent years
bootstrap percolation has grown beyond the field of disordered magnetism,
and found several new applications in physics, materials science, biology,
electronic communication, and social networks. It has also acquired a new
name and is sometimes called k-core percolation. We shall not go into
these aspects here, but an interested reader will find useful links to
these problems in reference~\cite{farrow2}.

Consider a lattice whose sites are either occupied with probability $p$ or
empty with probability $1-p$. The empty sites need not be devoid of matter
but rather they may be occupied by a different species of atoms that are
not relevant for our discussion. The relevant interactions are among pairs
of nearest neighbor occupied sites. In the limit $p \rightarrow 0$, the
occupied sites are isolated and can not exhibit cooperative effects. On
the other hand, they exhibit strong cooperative effects in the limit $p
\rightarrow 1$. Intuitively, we expect the transition from independent to
cooperative behavior at a threshold $p=p_c$ when the occupied sites begin
to form a connected cluster spanning the entire lattice. This is known as
the percolation point, and $p_c$ the percolation
probability~\cite{fisher}. The usual approach of percolation theory is to
start with a lattice with a fraction $p$ of its sites occupied randomly,
and search for the smallest value of $p$ where the occupied sites contain
a spanning cluster with probability unity. Bootstrap
percolation~\cite{chalupa} asks a slightly more general question. Given a
lattice (of coordination number $z$) whose sites are randomly occupied
with probability $p$, what is the smallest value of $p$ at which the
lattice contains a spanning cluster of occupied sites such that each site
in the cluster has at least $m$ ($0 \le m \le z$)  nearest neighbors. The
usual approach to this problem is again to start with a lattice with a
fraction $p$ of its sites occupied randomly, and recursively cull all
sites that have less than $m$ neighbors. The smallest value of $p$ for
which a spanning cluster survives the culling process is the $m$-bootstrap
percolation probability. The bootstrap percolation probability for $m=2$
is identical with the ordinary percolation probability. Thus the
$m$-bootstrap process is a generalization of the ordinary percolation
process. In the following, we shall use the word percolation to mean
bootstrap percolation for an arbitrary value of $m$ unless specified
otherwise.

In order to determine $p_c$ numerically, one has to examine a randomly 
occupied lattice over a range of values of $p$. For each $p$, a 
configuration of occupied sites is generated. Then sites with less than 
$m$ nearest neighbors are culled recursively. This procedure either 
empties the lattice, or yields a stable configuration (called an $m$-core) 
in which every occupied site has at least $m$ occupied nearest neighbors. 
The data is averaged over several configurations for each value of $p$. 
The smallest value of $p$ at which the probability of finding an $m$-core 
is unity is identified with the percolation probability $p_c$. We call 
this a one-shot bootstrap procedure to emphasize that the initial 
configuration is subjected to bootstrap culling only once, and no stable 
site is culled in this procedure.

We now describe a different bootstrap process~\cite{manna,farrow1,farrow2} 
that we call the sequential bootstrap process. It comprises the following 
steps: \\ \noindent Step-1: Start with a fully occupied lattice.\\ Step-2: 
Attack an occupied site randomly i.e. remove it. \\ Step-3: Remove all 
unstable sites recursively, and count the total number of sites removed 
including the stable site removed in step-2.\\ Step-4: If the lattice is 
not empty, go to step-2 and repeat.

The result of the above exercise is a list of integers $\{a_1, a_2, a_3, 
\ldots \}$; where $a_k$ is the size of the $k$-th avalanche. Some 
interesting questions are: (i) how many attacks does it take to empty the 
lattice? If it takes $kmax$ attacks to empty a lattice of $N$ sites, the 
quantity of thermodynamic interest is the ratio $kmax/N$, (ii) what is the 
distribution of the avalanches $\{a_k\}$ ? (iii) the fractional occupation 
of the stable lattice ($p_c$) before the last attack, (iv) the size of the 
last avalanche $a_{kmax}$ that empties the lattice, and (v) the nature of 
fluctuations in the system just before the last avalanche. The reason for 
the interest in the size of the last avalanche is the following. If the 
last avalanche is microscopic, the transition at $p=p_c$ is second order. 
If the last avalanche is macroscopic, the transition is first order. 
Second order phase transitions exhibit critical fluctuations in their 
vicinity, and we find this to be the case here as well. Normally, first 
order transitions are not associated with critical fluctuations. However, 
in the present case, we find power-law (critical) fluctuations at the 
threshold of the first order transition ($p \rightarrow p_c^{+}$). In this 
respect, the bootstrap process is interesting. It provides a caricature of 
fluctuation driven first-order phase transitions.

Although it is not immediately obvious, the sequential bootstrap process 
described above has a close relationship with the one-shot bootstrap. 
Indeed, the results of the one-shot bootstrap can be obtained from those 
of the sequential bootstrap. Thus the sequential bootstrap is a more 
general bootstrap process, and includes one-shot bootstrap as a special 
case. We shall return to this point after we have established some results 
concerning the one-shot process on a Cayley tree.

\section{Bootstrap dynamics on a Cayley tree}

An exact solution of the one-shot bootstrap dynamics on the Bethe lattice 
has been obtained ~\cite{farrow2} following the method developed in the 
context of the zero-temperature dynamics of random field Ising 
model~\cite{dhar,sabha}. It is useful to look at this analysis in somewhat 
more detail. We take a random configuration, and put culling tags on sites 
that have less than $m$ nearest neighbors. When a site is culled, it does 
not affect other culling tags. This means that the culling process is 
Abelian, and the final result does not depend on the order in which we 
cull unstable sites. We take advantage of this Abelian property, and 
initiate the culling process from the surface of the Cayley tree. Let the 
surface be $n+1$ steps away from the root of the tree, and $P^{n}(z,m,p)$ 
be the conditional probability that a site $n$ steps away from the root is 
not culled given that it has a nearest neighbor at a distance of $n-1$ 
steps from the root. In the following, the arguments in $P^{n}(z,m,p)$ 
will be omitted if doing so does not cause any confusion. Thus, $P^{n}$ is 
determined by the recursion relation,

\bea P^{n}= p \sum_{k=0}^{z-1} {{z-1}\choose k} 
[P^{n+1}]^{k}[1-P^{n+1}]^{z-1-k} p_{k+1}, \nonumber \\ 
p_{k+1}= 1 \mbox{ if }k+1 \ge m, \nonumber \\ 
p_{k+1}= 0 \mbox{ if }k+1 < m. \eea

The rationale for the above equation is as follows. A site at a distance 
of $n$ steps from the root has one nearest neighbor at level $n-1$ and 
$z-1$ nearest neighbors at level $n+1$. The neighbors at level $n+1$ are 
independent of each other in deciding whether to cull the site at level 
$n$. We iterate equation (1) starting at the boundary of a large tree with 
$P^{n+1}=p$, and work inward. The equation iterates to a fixed point 
solution $P^{*}$ in the limit $n \rightarrow \infty$; $P^{*}$ is the 
conditional probability that deep inside the tree, a nearest neighbor of 
the root is occupied given that the root is occupied. The unconditional 
probability that the root is occupied is given by, \be P = p 
\sum_{k=0}^{z} {z\choose k} [P^{*}]^{k}[1-P^{*}]^{z-k} p_{k} \ee

If the root is occupied it must lie on an unbroken network of occupied 
sites going all the way to the boundary of the tree. Thus $P$ is also the 
$m$-bootstrap percolation probability, i.e.  the probability that a site 
in the deep interior of the tree is a part of an infinite $m$-cluster. The 
equations apply to the deep interior of the tree because a stable fixed 
point of the recursion relation is insensitive to the boundary of the 
tree. The deep interior part of the Cayley tree is also known as the Bethe 
lattice, or a random graph of a fixed coordination number $z$. Equation 
(2) was also obtained\cite{chalupa} from a mean field argument that 
happens to be exact for the bootstrap problem on a Bethe lattice. Our 
approach emphasizes the Abelian property of the bootstrap dynamics and 
provides a rigorous justification for the mean field equations in this 
case. This is important because it is in general not possible to justify 
the mean field equations even on a Bethe lattice beyond the statement that 
they are self-consistent.

In the case $m=2$, i.e. $p_0=p_1=0$, and $p_2=p_3=\ldots=p_z=1$, equations 
(1) and (2) reduce to the well known equations for ordinary percolation on 
a Bethe lattice \cite{fisher}: \bea P^{*}=p[1-(1-P^*)^{z-1}], \nonumber \\ 
P=p[1-(1-P^*)^{z}- z P^*(1-P^*)^{z-1}] \quad (m=2)  \eea The ordinary 
percolation ($m=2$) is a second order transition. The percolation 
probability is given by $p_c=\frac{1}{(z-1)}$; the order parameter $P=0$ 
if $p \le p_c$, and increases continuously with $p$ for $p>p_c$.

For $m \ge 3$, the percolation transition on the Bethe lattice is a first 
order transition; $p_c$ as well as the jump in the order parameter $P$ at 
$p=p_c$ increases steadily with increasing values of $m$.  For $m=z$ 
($p_0=p_1=\ldots=p_{z-1}=0; p_z=1$), $P^*$ is determined by the equation 
$P^*=p[P^*]^{z-1}$. The possible solutions are $P^*=0$ and 
$P^*=p^{-\frac{1}{z-2}}$. But physically acceptable solutions are: $P^*=0$ 
for $p<1$, and $P^*=P=1$ for $p=1$, indicating a first order transition at 
$p_c=1$.

The case $m=z-1$ ($p_0=p_1=\ldots=p_{z-2}=0; p_{z-1}=p_z=1$) yields 
pleasing closed form expressions for $p_c$ and the discontinuity in the 
order parameter at $p=p_c$. In this case, $P^*$ is determined by the 
equation: \bea P^*=p[(z-1)(P^*)^{z-2}(1-P^*)+(P^*)^{z-1}] \quad (m=z-1) 
\eea Obviously $P^*=0$ is a solution of equation (4). Other non-zero 
solutions of equation (4) are given by the equation, \bea 
f(p,P^*)=p[(z-1)(P^*)^{z-3}(1-P^*)+(P^*)^{z-2}]-1 =0 \quad (m=z-1) \eea In 
the range $p_c < p \le 1$, equation (5) has two real solutions that merge 
at $p=p_c$ ($p_c$ to be calculated below), and vanish from the real axis 
for $p < p_c$. $P^*$ makes a discontinuous jump from $P^*=P^*_{disc}$ to 
$P^*=0$ as $p$ crosses the value $p=p_c$ from above. The non-zero solution 
for $P^*$ at the double root is determined by the equation, \bea \left. 
\frac{df(p_c,P^*)}{dP^*}\right\vert_{P^*=P^*_{disc}}=0 \Longrightarrow 
\left. P^*_{disc}=\frac{(z-1)(z-3)}{(z-2)^2} \right. \quad (m=z-1) \eea 
Substituting from equation (6) into equation (5) gives, \bea 
p_c=\left(\frac{z-3}{z-2}\right)\left[\frac{(z-2)^2}{(z-1)(z-3)} 
\right]^{z-2} \quad (m=z-1) \eea

\section{Power-law avalanches}

We now show that there are critical fluctuations in the size of the 
avalanche as $p \rightarrow p_c^{+}$, irrespective of whether there is a 
first order or a second order transition at $p=p_c$. Let us focus on a 
pair of nearest neighbor occupied sites in a stable $m$-core. Designate 
one of the occupied sites as the root of the tree. The other site lies at 
the vertex of a sub-tree connected to the root. Now remove the root by 
hand. Let $\pi_a$ be the probability that this causes an avalanche of 
size-$a$ on the sub-tree. For example, \be \pi_0=p \sum_{k=0}^{z-1} 
{z-1\choose k}[P^*]^{k}[1-P^*]^{z-1-k} p_{k}. \ee In general, it is 
convenient to work with a normalized generating function $\pi(x)$, \be 
\pi(x)=\sum_{a=0}^{\infty}\pi_a x^{a}; \quad [\pi(x=1)=P^*]. \ee The 
generating function is determined by the equation, \be \pi(x)=\pi_0
+ x p \sum_{k=0}^{z-1} {z-1\choose k} [\pi(x)]^{k} [1-P^*]^{z-1-k} 
 [p_{k+1}-p_{k}] \ee

When we remove the root, avalanches may be initiated on any one of the $z$ 
subtrees connected to the root if the vertex of that sub-tree is occupied. 
Let $\Pi_a$ be the probability that a total avalanche of size-$a$ is 
initiated by the removal of the root. The generating function for $\Pi_a$ 
and its equation is, \bea \left. \Pi(x)=\sum_{a=0}^{\infty}\Pi_a 
x^{a}\mbox{;  } \quad \quad \Pi_a= 
\frac{1}{a!}\frac{d^a\Pi(x)}{dx^a}\right\vert_{x=0}. \nonumber \\ \Pi(x)=x 
p \sum_{k=0}^{z} {z\choose k}[\pi(x)]^{k}[1-P^*]^{z-k} I(z,m,k) \eea The 
last factor $I(z,m,k)$ in the above equation takes care of the notion of 
"removing the root by hand".  It is the probability that a stable root 
with $k$ occupied nearest neighbors would become unstable if $p$ were to 
decrease by an infinitesimal amount from $p$ to $p-\delta p$.

We are primarily interested in the behavior of $\Pi_a$ in the limit $a 
\rightarrow \infty$. This is determined by $\Pi(x)$ in the limit $x 
\rightarrow 1$. We calculate $\Pi(x \rightarrow 1)$ for $z=4$, and $m=3$. 
The algebra is relatively simple in this case but the main result of the 
analysis holds for all $z \ge 4$ and $m \ge 3$. We note that $P^*(4,3,p)$ 
is either zero, or it is determined by the quadratic equation, \bea 2p 
[P^*]^2-3p P^* +1=0; \quad P^*=\left. \frac{3p \pm \sqrt{9p^2-8p}}{4p} 
\right.\quad (z=4,m=3) \eea The two non-zero solutions are real in the 
range $\frac{8}{9}\le p \le 1$, and merge into a double root 
$P^*_{disc}=\frac{3}{4}$ at $p_c=\frac{8}{9}$. For $p<p_c$, the solutions 
of the quadratic equation disappear from the real plane, and we are left 
with only one solution $P^*=0$ that is physically acceptable. Thus $P^*$ 
jumps from the value $P^*(4,3,p)=\frac{3}{4}$ to $P^*(4,3,p)=0$, as $p$ 
crosses the value $p_c=\frac{8}{9}$ from above. The order parameter $P(p)$ 
jumps from $P(p)=\frac{21}{32} \approx .65$ to $P(p)=0$. For $p > p_c$, we 
choose the upper sign of the square root. Writing $p=p_c+\delta p$, 
equation (12) gives to the leading order in $\delta p$, \be P^*(p_c+\delta 
p)=P^*_{disc}+ \frac{9}{8\sqrt{2}}{(\delta p)}^{\frac{1}{2}} \quad 
(z=4,m=3) \ee For $z=4$, $m=3$, equation (10) for $\pi(x)$ takes the 
quadratic form, \be 3 x p [1-P^*]\pi^2(x)-\pi(x)+p[P^*]^3=0 \quad 
(z=4,m=3) \ee Although equation (14) can be easily solved, it is 
instructive to examine it in the vicinity of $\pi(x=1)=P^*$. Let 
$x=1-\delta x$, and $\pi(x=1-\delta x)=P^*+\delta \pi$, where $\delta x$, 
and $\delta \pi$ are infinitesimally small quantities. To the leading 
order, we get \bea 3 p (1-P^*)[\delta \pi]^2 -\{1-6p(1-P^*)P^*\}{\delta 
\pi} \nonumber \\ -3p(1-P^*)(P^*)^2 \delta x =0 \quad (z=4,m=3) \eea The 
reason why we have included the second order term in $\delta \pi$ in the 
above equation is that the coefficient of the linear term in $\delta \pi$ 
vanishes at the percolation transition point $p=p_c$, $P^*=P^*_{disc}$. 
Thus, at the transition point $\delta \pi$ has a square root singularity 
in $\delta x$, i.e. \be \delta \pi(x) = P^*_{disc} (x-1)^{\frac{1}{2}}, 
\quad x \rightarrow 1, \ee . It follows from equation (11) that $\Pi(x)$ 
also has a square root singularity at $x=1$.\be \Pi(x) = \Pi(x=1) + 
C_{\Pi} (x-1)^{\frac{1}{2}}, \quad x \rightarrow 1 \ee where $C_{\Pi}$ is 
a constant. Using equation (11) and Stirling's formula, we get \be \Pi_a= 
C_{\Pi} \frac{1}{a!}\frac{(2a-3)!!}{2^a}= C_{\Pi} \frac{(2a)!}{a 
2^{2a}{a!}^2} \sim \frac{1}{a^{\frac{3}{2}}}; a \rightarrow \infty, p=p_c 
\ee Equations (16), (17), and (18) hold at the bootstrap transition point 
$p=p_c$. If $p=p_c+\delta p$ where $\delta p$ is infinitesimally small, 
the coefficient of the linear term in equation (15) becomes of the order 
of $(\delta p)^{\frac{1}{2}}$ in view of equation (13). In this case 
$\delta \pi$ is determined by a quadratic equation of the form, \be 
(\delta \pi)^2 + b (\delta p)^{\frac{1}{2}}\delta \pi - c^2 \delta x=0 \ee 
where b, and c are constants. To the lowest order in $\delta p$ and 
$\delta x$, $\delta \pi$ is given by \be \delta \pi = \frac{1}{2} 
\left[\{b^2 (\delta p)+ 4 c^2 \delta x\}^{\frac{1}{2}}-b(\delta 
p)^{\frac{1}{2}}\right] \ee Thus $\pi(x)$, and therefore $\Pi(x)$ has 
leading square root singularity at $x=1-\frac{b^2(\delta p)}{4c^2}$. 
Consequently, $\Pi_a$ scales as, \be \Pi_a \sim \frac{1}{a^{\frac{3}{2}}} 
\left[1+\frac{b^2(p-p_c)}{4c^2}\right]^{-a}; \quad a \rightarrow \infty, p 
\rightarrow p_c \ee

We note that the power-law distribution of avalanches at $p=p_c^+$ results 
from the fact that the coefficient of the linear term $\delta \pi$ in 
equation (15) vanishes at $p=p_c$. The coefficient of $\delta \pi$ 
vanishes because the equation determining $P^*$ has a double root at 
$p=p_c$. This property is not incidental to the case $z=4$ and $m=3$. It 
holds for all values of $z \ge 4$, and all values of $m$ in the range $z 
\ge m \ge 3$. In experiments as well as computer simulations of the model, 
one measures the integrated probability distribution of avalanches 
$\Pi_a^{int}$, say an avalanche of size-$a$ in a range $p=p_1$ to $p=p_2$. 
If we are far from the transition point, the probability of large 
avalanches is exponentially small. Thus the integrated probability of 
large avalanches is mainly determined by the fact whether the interval of 
integration includes the transition point or not. In case the interval 
contains the transition point, only a small region of of width 
$\frac{c^2}{b^2a}$ contributes significantly to the integral for large 
$a$, giving us \be \Pi_a^{int} \sim a^{-\frac{5}{2}}, \quad a \rightarrow 
\infty \ee

\section{One-shot vs. sequential bootstrap}

The one-shot bootstrap process yields the percolation probability P($p$) 
for a random initial configuration when each site is independently 
occupied with probability $p$. These configurations are almost always 
unstable under the $m$-bootstrap process, and the number of unstable sites 
in them is large. Therefore the size of the avalanche that reduces the 
initial configuration to a stable $m$-core in the one-shot bootstrap 
process is large. In contrast to this, the sequential bootstrap starts 
with a fully occupied lattice and reduces it to an empty lattice by a 
series of attacks and avalanches. The avalanches in sequential bootstrap 
connect two nearby stable $m$-cores. We have proved in the previous 
section that an avalanche caused by the removal of a site in a stable 
$m$-core is exponentially small unless the lattice is at its critical 
point. Thus the sequential percolation probability $P^{seq}(p)$ decreases 
linearly with lattice occupation probability $p$. As there are no loops on 
the Bethe lattice, any stable $m$-core is necessarily infinite, and the 
probability $p$ that a site is occupied is identical with the probability 
$P^{seq}(p)$ that it lies on an infinite cluster.

In order to illustrate the connection between one-shot and sequential 
bootstrap we obtain the one-shot percolation probability P($p$) from the 
data of the sequential process. The key point linking the two processes is 
the following. Imagine we start with a fully occupied lattice, attack a 
fraction $p$ of the occupied sites sequentially (without accompanying 
avalanches) and then subject the lattice to a bootstrap culling procedure. 
This will of course give the same result as the one-shot process. The 
sequential process allows avalanches between successive attacks, so 
several sites that would have been attacked in the hypothetical situation 
imagined above are not available to be attacked because they have been 
culled in avalanches. Let us call such incidences false attacks. In order 
to get the one-shot result from the sequential process, we should keep 
track of the number $k$ actual attacks in the sequential process, as well 
as the number of false attacks up to the $k$-th attack. When the fraction 
of actual attacks plus the fraction of false attacks equals $p$ the 
sequential percolation probability $P^{seq}(p)$ would equal the 
percolation probability P($p$) under the one-shot process. This is shown 
in Figure-1 for $m=2$ (second order transition) and $m=3$ (first order 
transition) on a random graph with $z$=4.

\section{Avalanche distributions and order of transition}

We have shown that the cumulative probability of a large avalanche of size 
$a$ on a random graph decays as $a^{-5/2}$ in the case of 1st as well as 
2nd order percolation transitions. In Figure-2 we have plotted the 
cumulative probability distribution C($a$) vs. $a$ on a 4-coordinated 
random graph for $m=2$ and $m=3$. In order to obtain good quality data in 
reasonable computer time (one day), we performed the simulations on a 
graph of $10^4$ nodes; the data was averaged over $10^2$ independent 
assignments of nearest neighbors for each node (wirings), and $10^3$ runs 
of the sequential bootstrap process for each wiring. Thus the average is 
taken over $10^5$ samples of a $10^4$-node network. Probability is 
calculated as probability per node. Hence C($a$) has a range from 
$10^{-9}$ to 1. The avalanche is normalized by dividing with the total 
number of nodes so that $a$ varies from $10^{-4}$ to 1.

In the case of a 1st order transition, Figure-2 shows that avalanches fall 
into two disjointed categories. A category of extensive avalanches that 
are of the order of the size of the network (1st order jump 
discontinuity), and a category of microscopic avalanches preceding the 1st 
order jump. The microscopic avalanches show a remarkable $a^{-5/2}$ power 
law over several decades even in a relatively small network of $10^4$ 
nodes. In contrast to this, the power law behavior of C($a$) in the case 
of the 2nd order transition is rather difficult to see because it is 
masked by strong cutoff effects. At first sight, this is surprising 
because our analysis predicts a power-law behavior at 1st as well as 2nd 
order transition. It is common to see power-law behavior at a 2nd order 
transition rather than a 1st order transition. The present simulations 
appear to show the opposite of what we commonly encounter. This unexpected 
behavior is not confined to random graphs, but is also seen on periodic 
lattices and small word networks~\cite{farrow2}. Indeed, it may be used to 
identify a 1st order transition that is otherwise more difficult to decide 
numerically~\cite{parisi}.

The power-law distribution of avalanches at the percolation point applies 
in the thermodynamic limit. A finite system will always have a cutoff. 
Therefore the question is why the cutoff is so much shorter in the case of 
a 2nd order transition? In other words, why the largest avalanche at a 2nd 
order transition is much shorter than the largest avalanche at a 1st order 
transition? The answer is relatively easy on a random graph, and it may 
apply to other lattices as well. On a random graph or equivalently on a 
Bethe lattice, finite (non-spanning)  $m$-cores are unstable under 
bootstrap dynamics for $m \ge 2$. An occupied site must necessarily belong 
to a spanning cluster. However, the number of sites in a spanning cluster 
may vary over a very wide range covering all the sites on the lattice to a 
relatively small number of sites that suffice to make a spanning path 
through the system. For example, on a Cayley tree of $n$ generations, the 
total number of sites is of the order of $N \approx (z-1)^n$, but it takes 
only 2$n$ steps to go from one point on the surface to another point on 
the surface through the root of the tree. At the 2nd order percolation 
transition the system will fluctuate between an empty lattice and one 
having approximately $\log_{z-1} N$ occupied sites. In contrast to this, 
at a 1st order transition the system will fluctuate between an empty 
lattice and one containing occupied sites of the order of $N$. This may 
explain why the cutoff scales as $N$ at a 1st order transition, and as 
$\log_{z-1} N$ at a second order transition. Largest avalanches scaling as 
$N$ at 1st order transition and $\log_{z-1} N$ at 2nd order transition are 
also observed on hypercubic, bcc, and triangular lattices as well as small 
world networks. This suggests that finite $m$-cores on periodic lattices 
may not contribute to an increase in the cutoff of the scale of 
avalanches. This is also indicated by the data shown in Figure-2 for the 
case $m$=1; there are finite clusters in a stable 1-core on a Bethe 
lattice, but its avalanche distribution is characterized by a shorter 
cutoff in comparison with the case $m$=2. However, more work and thought 
is needed to fully understand these aspects of the problem.

\begin{figure}[p]
\includegraphics[width=.7\textwidth,angle=-90]{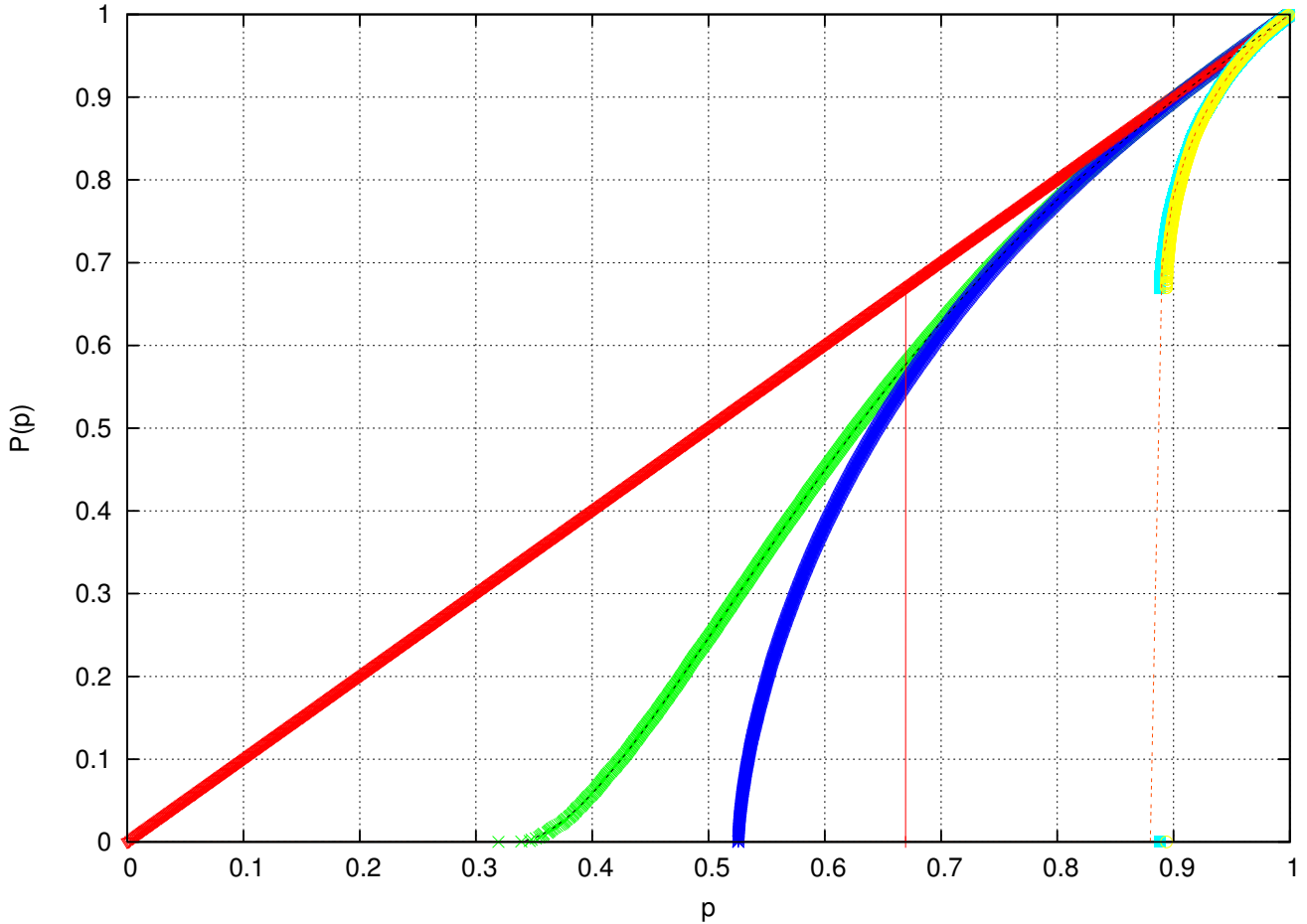}

\caption{The probability P($p$) that a site on a 4-coordinated random 
graph belongs to a spanning $m$-core under sequential bootstrap dynamics. 
Curves should be viewed starting from the top right corner (fully occupied 
lattice). The x-axis has a different meaning for each curve. Two nearly 
straight and overlapping lines that terminate at $p \approx .66$ and $p 
\approx 0$ respectively correspond to 1st ($m$=3) and 2nd ($m=2$)  order 
transitions: in their case the x-axis denotes fractional occupation of the 
input lattice. The lines are nearly straight because the avalanches along 
them are exponentially small. The remaining curves are based the same data 
but plotted against different variables. For the two curves in the middle 
that terminate at $p \approx .52$ and $p \approx .33$; as well as the two 
short curves on the right that terminate at $p \approx .88$ and $P(p) 
\approx .65$; the x-axis denotes the 1-$f_a$ and $1-f_t$ respectively 
where $f_a$ is the fractional number of sequential attacks, and $f_t$ is 
the fractional number of total attacks including false attacks (see text). 
The two curves in the middle are for $m=2$, and on the right for $m=3$. 
The curves terminating at $p=.33$ and $p=.88$ are identical with one-shot 
bootstrap percolation probability $P(p)$ for $m=2$ and $m=3$ respectively, 
as shown by the corresponding theoretical result superimposed upon them.}

\label{fig1}

\end{figure}

\begin{figure}[p]
\includegraphics[width=.7\textwidth,angle=-90]{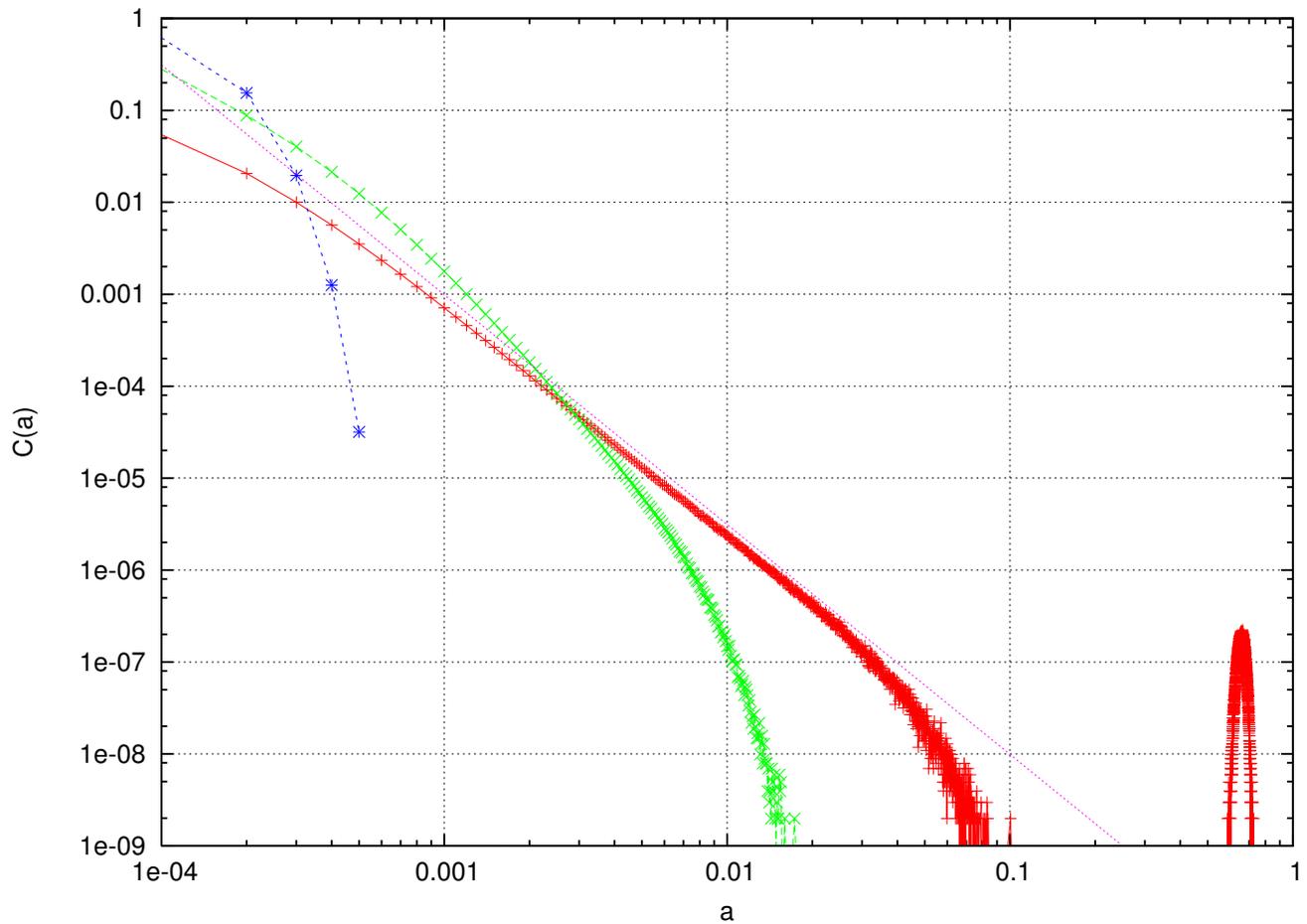}

\caption{ Distribution of cumulative avalanches in a sequential bootstrap 
process for $m$=1 (blue curve), $m$=2 (green curve), and $m$=3 (red 
curve). The cases $m$=1 and $m$=2 correspond to continuous transitions; 
these cases show truncated avalanches. A discontinuous transition ($m$=3) 
shows a striking power-law over several decades. A line with slope $-5/2$ 
is shown for comparison. This effect is seen in periodic lattices as well 
(see text). In addition, avalanche distribution in a discontinuous 
transition is characterized by an isolated sharp peak corresponding to 
extensive avalanches that empty the lattice.}

\label{fig2}

\end{figure}

\end{document}